\documentclass[trackchanges]{aastex62}

\usepackage{CJK}
\usepackage{amsmath}
\usepackage{amssymb}
\usepackage{tikz}
\usetikzlibrary{arrows}
\usepackage{booktabs}
\usepackage{multirow}
\usepackage{tabularx}
\usepackage{booktabs,multirow}

\shorttitle{Failed filament eruptions}
\shortauthors{Zhou et al.}
\begin{document}
\begin{CJK*}{UTF8}{gbsn}

\title{Why torus-unstable solar filaments experience failed eruption?}

\correspondingauthor{Zhenjun Zhou, Xin Cheng}
\email{zhouzhj7@mail.sysu.edu.cn, xincheng@nju.edu.cn}

\author[0000-0001-7276-3208]{Zhenjun Zhou(周振军)}
\affiliation{School of Atmospheric Sciences, Sun Yat-sen University, Zhuhai, Guangdong, 519000, China}
\affiliation{CAS Key Laboratory of Solar Activity, National Astronomical Observatories, Beijing 100012, China}
\affiliation{CAS Center for Excellence in Comparative Planetology, China}
\affiliation{Southern Marine Science and Engineering Guangdong Laboratory (Zhuhai)}
\author[0000-0003-2837-7136]{Xin Cheng}
\affiliation{School of Astronomy and Space Science, Nanjing University, Nanjing 210093, People's Republic of China}
\affiliation{Max Planck Institute for Solar System Research, Gottingen, 37077, Germany}
\author[0000-0003-0951-2486]{Jie Zhang}
\affiliation{Department of Physics and Astronomy, George Mason University, Fairfax, VA 22030, USA}
\author[0000-0002-8887-3919]{Yuming Wang}
\affiliation{CAS Key Laboratory of Geospace Environment, Department of Geophysics and Planetary Sciences, University of Science and Technology of China, Hefei, Anhui 230026, China}
\affiliation{CAS Center for Excellence in Comparative Planetology, China}
\author[0000-0001-7548-0051]{Dong Wang}
\affiliation{Department of Mathematics and Physics, Anhui Jianzhu University, Hefei, Anhui 230601, People's Republic of China}
\affiliation{CAS Key Laboratory of Geospace Environment, Department of Geophysics and Planetary Sciences, University of Science and Technology of China, Hefei, Anhui 230026, China}
\author[0000-0001-6804-848X]{Lijuan Liu}
\affiliation{School of Atmospheric Sciences, Sun Yat-sen University, Zhuhai, Guangdong, 519000, China}
\affiliation{CAS Center for Excellence in Comparative Planetology, China}
\author[0000-0002-5996-0693]{Bin Zhuang}
\affiliation{CAS Key Laboratory of Geospace Environment, Department of Geophysics and Planetary Sciences, University of Science and Technology of China, Hefei, Anhui 230026, China}
\author[0000-0002-4721-8184]{Jun Cui}
\affiliation{School of Atmospheric Sciences, Sun Yat-sen University, Zhuhai, Guangdong, 519000, China}
\affiliation{CAS Center for Excellence in Comparative Planetology, China}

\begin{abstract}
To investigate the factors that control the success and/or failure of solar eruptions, we study the magnetic field and 3-Dimensional (3D) configuration of 16 filament eruptions during 2010 July - 2013 February. All these events, i.e., erupted but failed to be ejected to become a coronal mass ejection (CME), are failed eruptions with the filament maximum height exceeding $100 Mm$. The magnetic field of filament source regions is approximated by a potential field extrapolation method. The filament 3D configuration is reconstructed from three vantage points by the observations of STEREO Ahead/Behind and SDO spacecraft. We calculate the decay index at the apex of these failed filaments and find that in 7 cases, their apex decay indexes exceed the theoretical threshold ($n_{crit} = 1.5$) of the torus instability.  We further determine the orientation change or rotation angle of each filament top during the eruption.  Finally, the distribution of these events in the parameter space of rotation angle versus decay index is established. Four distinct regimes in the parameter space are empirically identified. We find that, all the torus-unstable cases (decay index $n > 1.5$), have a large rotation angles ranging from $50^\circ - 130^\circ$. The possible mechanisms leading to the rotation and failed eruption are discussed. These results imply that, besides the torus instability, the rotation motion during the eruption may also play a significant role in solar eruptions.

\end{abstract}
\keywords{Sun: corona --- Sun: filaments, prominences --- Sun: coronal mass ejections (CMEs) --- instabilities}

\section{Introduction} \label{sec:intro}
Coronal Mass Ejections, or CMEs, are the most spectacular burst of plasma and magnetic field in the Sun's corona. They are frequently associated with solar flares. CME and flare are considered to be two observational aspects of the same physical process in a solar eruption \citep{Harrison_1996,Zhang_etal_2001,Zhang_etal_2004,priest_forbes_2002}.  

Magnetic flux ropes (MFRs, a set of coiled magnetic field lines winding more than once about a common axis) are believed to be the fundamental structure of CMEs. Coronagraph  images of CMEs and in situ measurements of magnetic field validate that the MFR configuration of CMEs does exist post the solar eruption \citep{Burlaga_etal_1981,Vourlidas_etal_2013}. However it is still debated whether an MFR is present in the corona prior to an eruption or is formed during the eruption process. Some observational features could contain hints as closely related to the MFRs, which include filaments, sigmoids, and hot channels \citep{Kuperus_Raadu_1974,Rust_Kumar_1994,McKenzie_Canfield_2008,Zhang_etal_2012,Cheng_etal_2013}; these features may be just different manifestations of MFRs, depending on different observational selection effect (e.g., sensitive to different temperatures), perspectives, as well as magnetic environment \citep{Cheng_etal_2017}.
Filaments are known to be made of cold and dense plasma suspended in the magnetic dips of an MFR configuration \citep{Mackay_etal_2010,Guo_etal_2010}. Filaments are therefore a good tracer of MFRs in the corona \citep{Schmieder_etal_2013,Zhou_etal_2017}. 

However, the MFR eruptions are not always associated with CMEs. For a so-called ``failed'' filament eruption, a strong deceleration appears in the wake of the initially eruptive-like acceleration, the eruptive filament reaches a maximum height as the mass in the filament threads drains back toward the Sun \citep{Ji_etal_2003} and no propagating CME in the white-light coronagraph images. The popular belief attribute such failure to the criteria for the torus instability (TI, in general terms: a sufficiently steep decrease of the overlying field with height) is not met at or above the eruption site \citep[e.g.,][]{Torok_Kliem_2005,Kliem_Torok_2006,Liu_2008,Liu_etal_2012,Song_etal_2014}.
The critical value is generally suggested to be typically in a range of 1.1-1.5 \citep[e.g.,][]{Kliem_Torok_2006,Demoulin_Aulanier_2010,Olmedo_Zhang_2010,Zuccarello_etal_2015}.
Some filament eruptions exhibit a strong rotation motion about its ascending direction and display a characteristic ``inverse $\gamma$'' shape, which refers to as the Kink instability \citep[e.g.,][]{Hood_Priest_1979,Torok_Kliem_2005}. However, kink instability is not an effective mechanism for full solar eruptions. It is often needs to cooperate with a torus instability \citep[e.g.,][]{Kliem_Torok_2006,Liu_2008,Schmieder_etal_2013,Vemareddy_Zhang_2014}.  

Recently, an experimental result demonstrates that torus-driven eruptions can fail under weak kink onset condition \citep{Myers_etal_2015}. Using solar observations, \citet{Jing_etal_2018} pointed out that the TI onset criteria is not a necessary condition for CMEs, some TI-stable MFRs can manage to break through the strong ``strapping'' field and evolve into CMEs. The eruption is additionally influenced by other factors such as the ${{ \mathcal T }}_{w}$ (twist number in the MFRs; \citealt{Myers_etal_2015,Liu_etal_2016}), $\Delta \varphi$ (the change of orientation of the polarity inversion line (PIL) as a function of height; \citealt{Baumgartner_etal_2018}). 
Meanwhile, with a strong writhing, the erupting MFR may experience a dissolution by magnetic reconnection with the overlying flux, resulting a failed eruption \citep{Hassanin_Kliem_2016}.
Anyway, most of the previous observational studies of failed eruption could not reveal the exact mechanism associated with it.

Uncovering what prevents an evolving eruption from becoming ejective surely improves 
our understanding of the requirements for a solar eruption. Using the 3D reconstruction by exploiting observations of multiple views and the potential field source surface (PFSS) model \citep{schrijver_rosa_2003}, we have investigated 16 failed filament eruptions. We find out 
that the writhe of failed filament eruption varies significantly from event to event, and the amount of writhe depends on the decay index of strapping magnetic field. 
In Section~\ref{sec:Observation}, we describe our event sample as well as the data and methods used. The details of the analysis are described in Section~\ref{sec:Observation}, and the obtained results and discussions are presented in Section~\ref{sec:Results}.

\section{Observation \& Analysis}\label{sec:Observation}
\subsection{Instruments}\label{Instruments} 

The twin Solar TErrestrial RElations Observatory (STEREO) A (Ahead), B (Behind) and Solar Dynamics Observatory (SDO) provide us an unprecedented opportunity to observe filaments in a multi-view setting. The Atmospheric Imaging Assembly \citep[AIA;][]{lemen_etal_2012} on board SDO can observe a filament in narrow extreme-UV (EUV) passbands including 304 {\AA} (formation temperature
$T_f = 10^5 K$) and  193 {\AA} ($T_f = 1.58 \times 10^6 K$) with a high cadence (12 s), high spatial resolution ($0\farcs6$ per pixel), and large field of view (FOV; $1.3 R_\odot$).  Meanwhile the Extreme Ultraviolet Imager (EUVI) on board STEREO provides another view of the filament at similar wavelengths, i.e. 304 {\AA} ($T_f = 6 \sim 8 \times 10^5 K$) and 195 {\AA} ($T_f = 1.4 \times 10^6 K$) with
a FOV of $1.7 R_\odot$ \citep{Howard_etal_2008}. For a failed filament eruption, its evolutions of the height and velocity have exactly the same trend as the hot-channel prior to its ceases to rise \citep{Cheng_etal_2014}.  Utilizing these multi-view observations, we apply 3D reconstruction to obtain the 3D configuration and evolution of filaments of study. The Helioseismic and Magnetic Imager \citep[HMI;][]{Schou_etal_2012}, also on board SDO, provides photospheric vector magnetic field data with a cadence up to 45s and a pixel size of $0\farcs5$. We have employed three different coronagraphs, 
Solar and Heliospheric Observatory (SOHO)/Large Angle and Spectrometric Coronagraph (LASCO)-C2 \citep{Brueckner_etal_1995}, STEREO/Sun Earth Connection Coronal and Heliospheric Investigation (SECCHI)-COR1 A and B \citep{Howard_etal_2008}, to determine whether a filament eruption results in CME or not, i.e., a successful eruption or a failed eruption. 

\subsection{Selection of Events} \label{sec:Selection}

16 failed filament eruptions are selected in this study (Table~\ref{tab:mytable}) according to the following criteria (e.g., Figure~\ref{fig:figure1}):
 (1) It is a failed filament eruption, i.e., no corresponding CME is captured in LASCO/C2 or SECCHI/COR1 (figure~\ref{fig:figure1}(b));  
 (2) The source region of the filament should be located on the solar disk in the view of SDO/AIA to allow for the coronal magnetic field extrapolation, as well as in the limb view of STEREO/EUVI for a necessary of 3D reconstructions (e.g., Figure~\ref{fig:figure1}(c,d)); 
 (3) The terminal height of the filament can be exactly determined. In this study, we only consider the cases of which the maximum height exceeds 100 Mm.
An erupting filament that stops at a lower altitude is inclined to be torus-stable in its later evolution due to a ``relatively high probability'' of a small decay index at the lower heights. Since the purpose of this study is to examine the nature of failure of torus-unstable events, a choice of high heights makes our selection of event unambiguous. 

Based on these criteria, we examine SDO/AIA and STEREO/EUVI data to search for suitable filament eruption events from 2010 July to 2013 February, during which the near-quadrature configuration of  STEREO A/B allows for the best 3D view of a solar eruption (see Figure~\ref{fig:figure1}(a)).
We have successfully identified 16 such events, which are listed in Table~\ref{tab:mytable}.  Through browsing the evolution of these 16 filament eruptions, we find out part of these cases show a strong rotation motion, hence we focus on the relationship between the rotation motion and filament eruption.  
 
\subsection{Decay Index \& Rotation Angle} \label{sec:Method}
For the 16 selected events, we create a parameter space that characterizes the torus instability and the writhing morphological change.
The critical parameter for the torus instability is the decay index ($n=-d\ln B_{ex}/dln h$, where $B_{ex}$ is the horizontal component of external field perpendicular to the radial component $B_r$ in spherical coordinates). Here we employ the PFSS model to calculate the coronal magnetic field based on the synoptic map of the photospheric radial field.
It should be noted that, only the transverse component of the extrapolated potential field is used, since the radial component does not contribute to the downward confinement onto the erupting MFRs. The final decay index is an average value along the main PIL.
We use 2012 May 5 event (No.9 in table~\ref{tab:mytable}) as an example to demonstrate how the decay index at its maximum height is calculated. Figure~\ref{fig:figure2}(a) and (b) show the erupted filament stopping at a certain height in SDO and STEREO-B view angles. We reconstruct the 3D coordinates of several selected points along the erupted filament axis using scc\_measure.pro routine, which is available in SolarSoftWare \citep{Freeland_Handy_2012}.  The maximum height of the filament is thus determined to a good degree. We sample the segment of the PIL directly underneath the filament by clicking on the segment as uniformly as possible to get sufficient representative points (marked by cyan line in Figure~\ref{fig:figure2}(c)), and then calculate the decay index $n$ at different heights for each selected point. In Figure~ \ref{fig:figure2}(d), we plot $n$ as a function of $h$, which is averaged over all selected points, with the error bar indicating the standard deviation. 
The filament final decay index corresponding to the maximum height can be found through interpolation of these discrete $n(h)$ nodal values, the uncertainty of the final decay index can also be estimated by interpolation.  For this case, we obtain that the decay index at maximum height $n_{maxh}=2.20\pm0.09$. Note that the threshold value of torus instability is believed to be 1.5 for a toroidal current channel \citep{Kliem_Torok_2006}. Thus, this derived $n_{maxh}$ is significantly larger than the theoretical critical value. In the meanwhile, $n$ increases monotonically as the height increases, so there is no local torus-stable confinement \citep{Wang_etal_2017}. Obviously, this filament eruption is in the torus unstable state but failed.

Here, we look into the writhing morphological change during the eruption of these events. The writhe is proportional to the difference in angle between the tangent vector at the top and the line connecting the footpoints \citep{Torok_etal_2010}. To evaluate the writhe during the eruption, we calculate the rotation angle $\varphi$ from the reconstructed filament. The same case is employed as the example. We project the erupted filament onto the solar disk from the top view (See Figure~\ref{fig:figure3}(a)).  Here we use the line connecting the elbows as the proxy of the tangent vector at the top.  Four points (white asterisks in Figure~\ref{fig:figure3}(a-b)) selected near the two elbows are used for fitting. The projected filament top is represented by a fitted regression line. $\varphi$ is then given as the difference in angle between the fitted regression line and the line connecting the footpoints. 
The image sequence (Figure~\ref{fig:figure1}(d)) also shows that the rotation is of the sense of clockwise (CW) (viewed from above) of this filament eruption. For this case, we calculated the rotation angle and its corresponding error ($\varphi=130^{\circ}\pm1.6^{\circ}$). Its error originates from the uncertainty of the elbow's location.

   \begin{deluxetable}{ccccccccc}
   \tablenum{1}
   \tablecaption{Filament list}
   \label{tab:mytable}
   \tablewidth{0pt}
   \tablehead{
   \colhead{Number} & \colhead{Date} & \colhead{Time\tablenotemark{a}}&
   \multicolumn{2}{c}{Location} & \colhead{Flare} & \colhead{$h_\mathrm{max}$\tablenotemark{b}} &\colhead{$\varphi$\tablenotemark{c}} & \colhead{$n_\mathrm{maxh}$\tablenotemark{d}} \\
   \cline{4-5}
   \colhead{} & \colhead{YYYYMMDD} & \colhead{hhmm} & \colhead{Type} & \colhead{Position} & \colhead{} & \colhead{(Mm)} & \colhead{($^\circ$)} & \colhead{} 
   }
   \decimals
   \startdata
   1           & 20100722 & 2307 & QS   & N47W22 & *         & 126 & 10  & 0.63  \\
   2           & 20110724 & 0736 & QS   & N61W69 & *         & 329 & 2   & 1.02  \\
   3           & 20110728 & 0056 & QS   & N37E25 & *         & 167 & 54  & 0.99  \\
   4           & 20110928 & 0207 & QS   & N39E09 & *         & 147 & 114 & 1.49  \\
   5           & 20111104 & 1927 & AR   & N44E23 & *         & 209 & 83  & 0.69  \\
   6           & 20111225 & 1146 & QS   & S25W23 & C8.4      & 155 & 86  & 2.80  \\
   7           & 20120101 & 0137 & AR   & N24W34 & *         & 162 & 16  & 1.43  \\
   8           & 20120304 & 1745 & AR   & N14W40 & C3.3      & 235 & 50  & 1.91  \\
   9           & 20120505 & 1746 & AR   & N16E35 & C3.0      & 129 & 130 & 2.20  \\
   10          & 20120811 & 1656 & AR   & S19E18 & C2.0      & 134 & 103 & 2.87  \\
   11          & 20120816 & 1826 & AR   & S22W49 & B5.3      & 172 & 99  & 1.64  \\
   12          & 20121025 & 0436 & AR   & N16W48 & C2.6      & 148 & 67  & 2.25  \\
   13{$^\dag$} & 20121112 & 0430 & AR   & S24W17 & *         & 186 & 87  & 1.72  \\
   14{$^\dag$} & 20121129 & 1220 & AR   & N15E58 & C4.5 C5.8 & 136 & 73  & 1.48  \\
   15          & 20130204 & 0117 & QS   & S52E88 & *         & 137 & 3   & 0.92  \\
   16          & 20130207 & 0226 & QS   & S49W79 & *         & 177 & 1   & 0.85  \\
   \enddata
   \tablenotetext{a}{Time of filament reaching its maximum height in FOV of STEREO EUVI.}
   \tablenotetext{b}{Reconstructed maximum height of the filament.}
   \tablenotetext{c}{Rotation angle during the eruption.}
   \tablenotetext{d}{The corresponding decay index at filament's maximum height position.}
   $^\dag$ In this table, events with Dagger-shaped symbol are observed by AIA at 193 {\AA} and EUVI at 195 {\AA}, the rest are  seen at 304 {\AA} by AIA and EUVI.
   \end{deluxetable}

\section{Results and Discussion} \label{sec:Results}
Figure~\ref{fig:figure4} shows the scatter diagram of TI parameters $n$ versus rotation angle $\varphi$ (with estimated uncertainties) for the 16 failed filament eruptions. 
The failed events with decay index $n$ less than 1.5 (9 out of 16 cases) may be consistent with the present understanding of the torus instability.
In the torus instability model, an erupting filament can't evolve into a CME when its decay index haven't achieved the theoretical expectation ($n \geqslant n_{crit}=1.5$) \citep{Kliem_Torok_2006}. 
However, exceptions to this theory do exist. The decay indexes of the other 7 cases (red color events in Figure~\ref{fig:figure4}) exceed more than 1.5, but they don't result into CMEs. This result argues against this conception that the torus instability is a sufficient condition for a full eruption.
Here we call these exceptions as torus-unstable failed eruptions.
Interestingly, all these torus-unstable events show a strong rotation during the eruptions. Their rotation angles ($\varphi$)  exceed $40^{\circ}$ with an average value of $89^{\circ}$.
The critical rotation angle, $\varphi \sim 40^{\circ}$, discriminates best between those torus-stable and torus-unstable failed filament eruptions. There is no one single case located in the region of large decay index ($n \geqslant 1.5$) and small rotation angle ($\varphi \leqslant 40^{\circ}$) regime.
Thus four distinct regimes can be empirically identified in the parameter space as shown in Figure~\ref{fig:figure4}. 

\begin{figure}[ht!]
\plotone{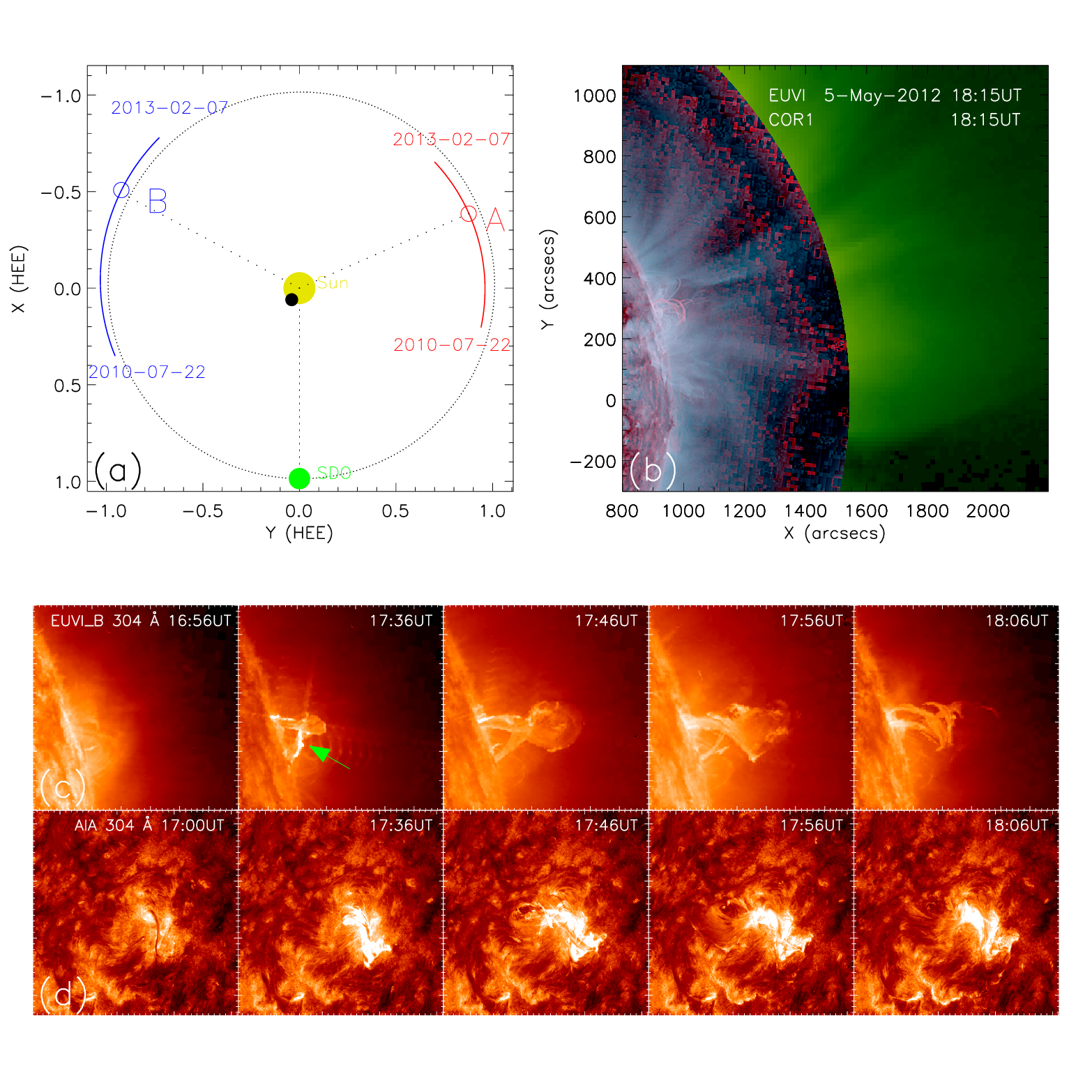}
\caption{The selection criteria of failed filaments. Panel (a) displays the paths of the STEREO-A (red arc) and B (blue arc) and position of SDO (green dot) in the ecliptic plane during the period from 2010 July 22 to 2013 February 07. The blue and red circles indicate the positions of STEREO-A/B on 2012 May 5 when a failed filament eruption occurred. The black dot on the Sun marks the filament source region, which appears on the solar disk when viewed from SDO, on the limb from STEREO-B, and on the backside of the Sun from STEREO-A.
Panel (b) shows no obvious CME signal in STEREO-B COR1 and EUVI 304 {\AA} and 195 {\AA} composite image acquired during the filament eruption.
Panels (c) and (d) provide observations of the prominence morphology during the eruption from the limb view in STEREO-A EUVI 304 {\AA} and disk view in SDO/AIA 304 {\AA}, respectively (The two-views of the eruption process in 304 {\AA} passband are available online as an animation).
\label{fig:figure1}}
\end{figure}

\begin{figure}[ht!]
\plotone{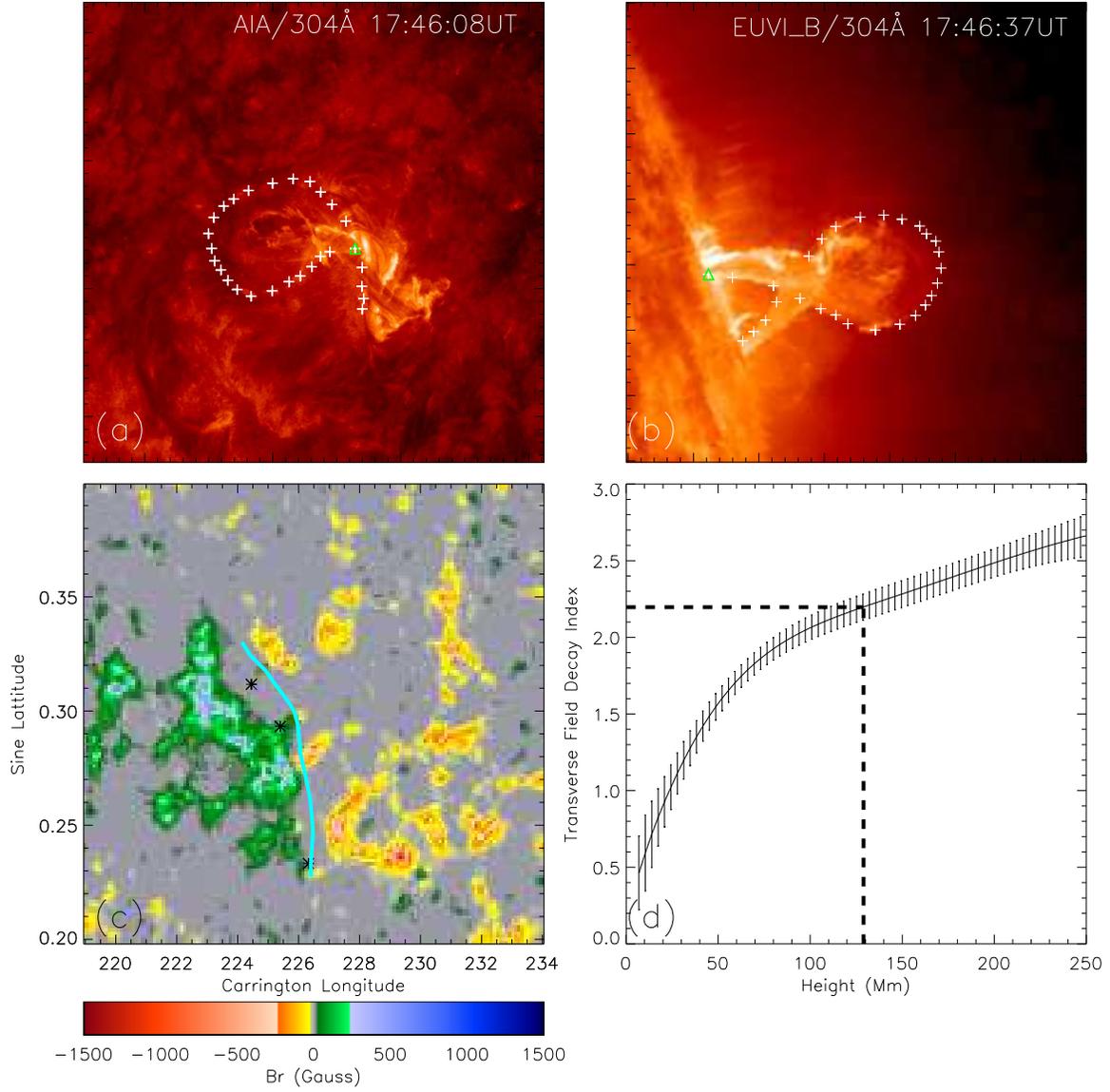}
\caption{Calculation of the decay index at the maximum height of the failed erupted filament. The white plus symbols in panels (a) and (b) depict the
prominence spine. The green triangle symbol denotes the same point viewed in two different angles (from SDO and STEREO-B). Panel (c) shows the line-of-sight magnetic field in the source region of the filament, black asterisks mark out the projected location of the filament before the eruption.  A cyan line denotes the PIL near this filament. In panel (d), the
decay index $n$ as a function of the height $h$ above the surface in units of Mm. The vertical and horizontal lines indicate the maximum height and the corresponding decay index.
\label{fig:figure2}}
\end{figure}

\begin{figure}[ht!]
\plotone{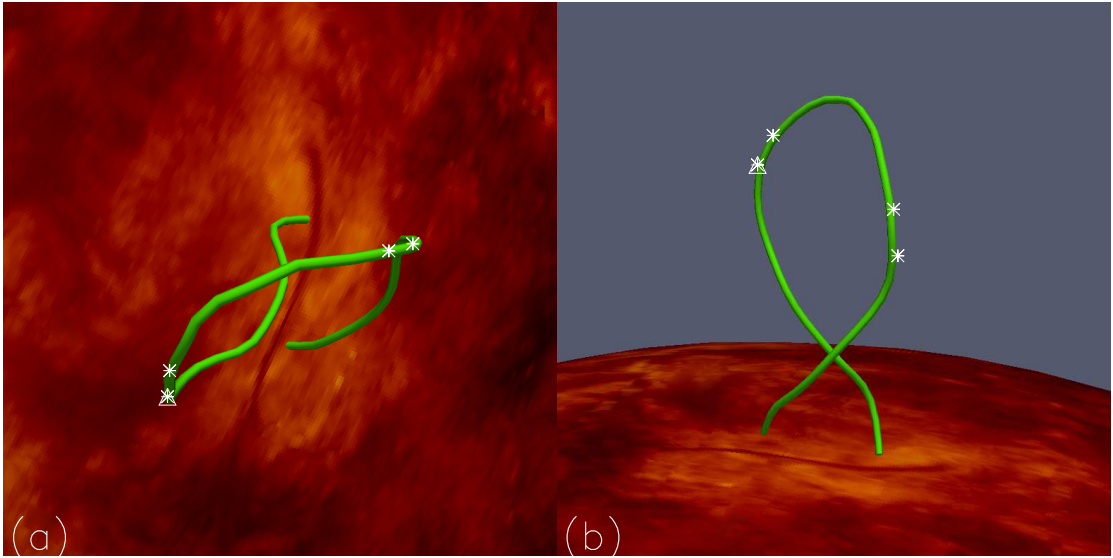}
\caption{Calculation of rotation angle during the eruption. The reconstructed 3D filament (colored in green) during the eruption from the top view (a) and side view (b), the bottom boundaries are the projected AIA 304 {\AA} synoptic map. Asterisks point to the shoulder of the filament. 
\label{fig:figure3}}
\end{figure}

\begin{figure}[ht!]
\plotone{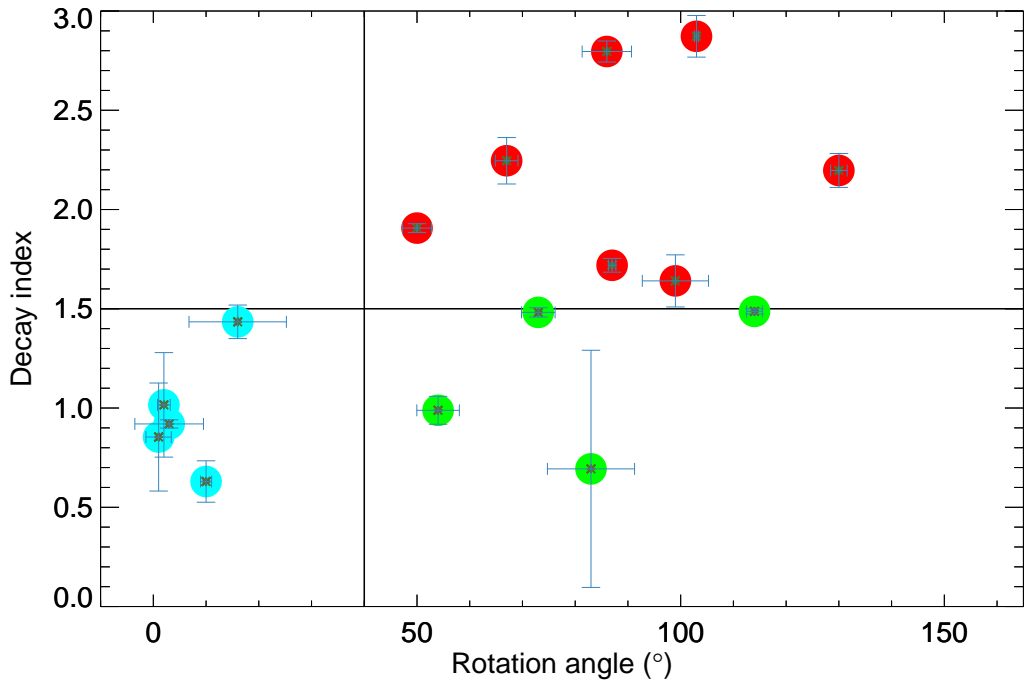}
\caption{Scatter diagram of rotation angle vs. decay index at the maximum heights for 16 failed filament eruptions. The vertical and horizontal black lines, which are empirically identified, delineate the four distinct instability parameter regimes described in the text. Cases in same regime are colored in same color. 
\label{fig:figure4}}
\end{figure}

Apparently, the rotation motion of a filament has a certain correlation with the failed eruption. 
Previous models concerning the writhing of MFRs have opposite effects for an eruption: 
On one hand, the writhing of the MFR's upper part into the orientation of the overlying arcade is energetically favourable for passing through the overlying arcade to become a CME \citep{Sturrock_etal_2001,Fan_2005}; 
On the other hand, the helical deformation facilitates interchange reconnection between filament flux and ambient flux \citep{Hassanin_Kliem_2016} and/or reconnection between the legs of the rope \citep{Alexander_etal_2006,Liu_Alexander_2009,Kliem_etal_2010}, such reconnection progressively decrease the flux content of the rope, up to its full destruction. 
This interaction is signified by the brightenings and non-thermal sources near the body or the crossing point of the filament \citep{Karlicky_Kliem_2010,Cheng_etal_2018}.
When only considering the torus-unstable failed eruptions, the reconnection caused by the MFR writhing seems dominant, an intense brightening in the body of the filament supports this possibility (See the brightening pointed by green arrow in 17:36 UT of figure~\ref{fig:figure1}(c)).
Simulation of \citet{Torok_etal_2010} pointed out that confined MFR eruptions tend to show stronger writhe at low heights than ejective eruptions (CMEs).
\citet{Hassanin_Kliem_2016} further inferred that if an eruption is halted, then the magnetic tension of the erupting flux can no longer be relaxed by expansion but only by further writhing, resulting in a tendency for confined eruptions to develop a strong writhing.

In summary, 16 failed filament eruptions are studied with both the AIA on board the SDO and EUVI on board the STEREO. Their decay indexes are obtained from the PFSS model and rotation angle are calculated with the help of the 3D reconstruction.  Thus we establish the scatter diagram of TI parameters $n$ versus rotation angle $\varphi$.
Seven cases are theoretically in torus-unstable state. Meanwhile, they all show strong writhing motions during the eruptions with rotation angle $\geqslant 40^{\circ}$.
It seems that writhing and failed eruption show a complex coupling relationship.
The possible reconnection due to the filament rotational motion may ruin the architecture of the MFR,
resulting a failed eruption. Simultaneously, this confinement induces a strong rotation instead of a further expansion.
More detailed observational analysis, theoretical considerations and numerical simulations are necessary towards a comprehensive understanding of the MFR eruption. 

\acknowledgments
We acknowledge the SECCHI, AIA and HMI consortia for providing the excellent observations. 
Z.J. is supported by the Open Research Program of CAS Key Laboratory of Solar Activity (KLSA201811).
X.C. is funded by NSFC grants 11722325,11733003, 11790303, 11790300, Jiangsu NSF grants BK20170011, and the Alexander von Humboldt foundation. 
J.Z. is supported by US NSF AGS-1249270 and NSF AGS-1156120.
D.W. acknowledges support by Natural Science Foundation of Anhui Province Education Department (KJ2017A493, gxyq2018030).
Y.W. is supported by NSFC grants 41574165, 41761134088 and 41774178. 
L.L. is supported by the Open Project of CAS Key Laboratory of Geospace Environment, and NSFC grants 11803096. 
J.C. is supported by NSFC grants 41525015 and 41774186.
Z.J. appreciates discussions and support with Prof. G.P.Zhou, Dr. C.Xia and Dr. Y.Guo. 
\software{SolarSoftWare \citep{Freeland_Handy_2012},
          Paraview \citep{Ahrens_etal_2005,Ayachit_Utkarsh_2015}}

\clearpage

 \end{CJK*}
\end{document}